# Visualization Techniques with Data Cubes

Utilizing Concurrency for Complex Data

Daniel Szelogowski

UW - Whitewater
Computer Science Master's Program
Whitewater, WI
szelogowdj19@uww.edu

**Abstract.** *With web and mobile platforms becoming more prominent devices utilized in data analysis, there are currently few systems which are not without flaw. In order to increase the performance of these systems and decrease errors of data oversimplification, we seek to understand how other programming languages can be used across these platforms which provide data and type safety, as well as utilizing concurrency to perform complex data manipulation tasks.*

## 1. INTRODUCTION

Visualization of complex data continues to be an issue as the size of our datasets grow larger and the platforms for these visualization tools expand alongside an increasingly mobile market. One approach, known as **Online Analytical Processing (OLAP)**, allows users to analyze complex information from multiple database systems simultaneously to "extract and view business data from different points of view" [1].

### 1.1 OLAP Systems and Data Cubes

The OLAP system divides databases into **data cubes (or OLAP Cubes)**, which allow for ease of analysis, reporting, and **visualization**. These cube objects allow for four basic operations:
- **Roll-up**
- **Drill-down**
- **Slice & dice**
- **Pivot (rotate)**

These functions allow us to manipulate the data cubes in particular ways that enable different viewpoints of the data sets. For example, the Roll-up function allows us to aggregate data and perform dimensional reduction - subsequently, the Drill-down function performs the opposite role, increasing a dimension; we may also Slice or Dice the data into sub-cubes of one or multiple dimensions, respectively, or Pivot the cube to view the data from a different rotation. As well, we can exploit these functions to create visualizations of our data through various platforms: 3D plane graphs, bar or line graphs, pivot tables, and various other charts or illustrations situationally.

### 1.2 Cube Issues

The ability to manipulate and aggregate data for visual analysis provided by OLAP cubes, while generally practical, features a series of flaws:

a) **Time:** Any modification to an OLAP cube requires a full update to the cube, which is time-consuming and resource-intensive.
b) **Oversimplification:** With visualization, it's very easy to take vastly complex data and simplify it to be more easily understandable; this may lead to assumptions on the data missing significant features, which may be highly inaccurate.
c) **Human Input:** While the OLAP system provides the tools for manipulating data to be visualized, human input is required for analysis operations, which may be intrinsically flawed. As such, an overreliance on visuals for quick, at-a-glance views can also become erroneous and worsen the problem of oversimplification.
d) **Visualization Democracy:** As data analysis grows more popular, tools for

various platforms and data sets are being rapidly developed, worsening both the issues of human input and oversimplification as we grow more reliant on visuals and fast (potentially non-critical) analysis.

**1.3 Present Research**
As such, this paper and subsequent research will seek to tackle the issues of time and oversimplification by means of concurrent algorithms to increase performance on varying platforms. In particular, we will look at the "CubesViewer" data visualization web application, since it is the most cross-compatible among multiple platforms.

The following sections will discuss current non-commercial dynamic systems and issues regarding their implementation, particularly on aspects of the programming languages involved, discussing the purpose and benefits of concurrency and more verbose languages, and how concurrency can be utilized to create a more efficient system of visualization.

## 2. OVERVIEW

Data cubes themselves present a few general pitfalls:
- Only using one cube or putting unrelated data in the same cube.
- Different levels of granularity between the dimension table and the fact table.
- Fact tables having more foreign key members than exist in the dimension table.

While these issues are difficult to avoid, current systems face difficulty in producing the visualization in general.

**2.1 Atoti**
The Python library "Atoti" solves issues of performance with Python visualization through the use of a double-backend system: a proprietary Java **interoperability (interop)** system (requiring either a pre-installed JDK or a fallback JDK bundled with the API) which performs the cube operation and subsequent OLAP functions, and a TypeScript JupyterLab extension which performs the visualization [3]. However, this implementation presents difficulties in reproducing the user-performed data analysis, as the operations and visualizations are performed within a live Jupyter notebook with the backend script which may be converted into a Python script using the 'nbconvert' command: a tool to convert Jupyter notebooks to formats such as PDF, HTML, LaTeX, Markdown, etc. [2].

Nonetheless, we see both an issue of shareability and platform compatibility; the dichotomy of 'fully client-side' and 'server-side with interop layers' is a problem for any data analyst seeking high compatibility, ease integration, little dependencies, and low overhead.

**2.2 CubesViewer**
Our primary focus will be on CubesViewer, an HTML5 application broken up as client (or "studio") and server models. The server model acts as a backend, serving the client application's web URLs using the Django library in Python; the client, more commonly referred to as the studio, performs the vast majority of its computations through JavaScript (JS), allowing easy integration into other websites and applications [4].

This implementation through JS comes at a cost, however, particularly in **performance**:
- JS is **single-threaded** by nature, albeit asynchronous, meaning we lose performance due to thread halting and decreased concurrency.
- JS is locked into memory constraints by the web-browser, rather than using something more modern like **WebAssembly (Wasm)** which can exploit desktop-grade performance in web applications and utilize features such as shared memory concurrency for parallel concurrency [5].
- JS has no support for 64-bit integers, so large, complex data sets may suffer in accuracy. As well, JS objects and

prototypes scale poorly with large applications, which may also be taxing on performance.

As such, while the ease of JS integration across platforms is admittedly useful, visualizing data while encountering these errors, similar to the ones faced by Python and the Atoti library, becomes a continuous battle for the aforementioned issues: **time** (program duration or resource cost) and **oversimplification of data**.

**2.3 Performance and Precision**
While a similar implementation of CubesViewer could be done utilizing Python both client- and server-side, we still face these issues:
   a) Python is inherently slow due to its implementation of dynamic programming and being an interpreted programming language.
   b) Complex numbers are difficult to work within Python. In order to avoid oversimplification of our data, we need to rely on third-party libraries such as NumPy to provide things such as 128-bit floating-point numbers [6]. While this is not a direct issue necessarily, this adds an additional layer of time and resource deprivation to large data sets.
   c) Python features a Global Interpreter Lock (GIL) which adds to the hindrance on performance — similar to JavaScript's single-threaded nature, this imposes numerous restrictions on threads; in particular, we cannot utilize multiple CPUs (**multi-core concurrency**) [7] — perhaps the biggest drawback in performance.

Noting these obstacles, we can look toward utilizing the same viewer "studio" of CubesViewer but performing the major data manipulation algorithms outside of JavaScript, instead using a language such as with WebAssembly.

### 3. MAIN METHODS

To create a more viable, performance-forward implementation we need to better understand the importance of **concurrency**. Concurrency is a form of **multithreaded programming** where two or more tasks overlap in execution, such as when multiple processes are assigned across CPU cores by the kernel, allowing for simultaneous, "concurrent" execution of the processes, or when new connections arrive before prior connections are complete and need to be immediately handled. **Parallelism** is, by nature, a special case of concurrency where two or more tasks begin at the same time. This type of execution will be our primary focus, as it is typically the more common form of concurrent programming.

**3.1 Unit Testing**
As a demonstration of the necessity for concurrency and why Python and JavaScript are insufficient for very large datasets, the following algorithm was performed across both languages:

```
runtimes = []

for lcv in range(0, 1000) {
  myNumbers = []
  for i in range(0, 500000) {
    myNumbers.append(
      random.int(0, 100000)) )
  }

  start_time = time.now()
  quickSort(myNumbers)
  duration = time.now() - start_time
  runtimes.append(duration)
}

avgRun = sum(runtimes)/len(runtimes)
print("Average runtime: {avgRun} ms")
```

This gives us a general idea of the performance of the two languages with large data sets, as seen in Figure 1.

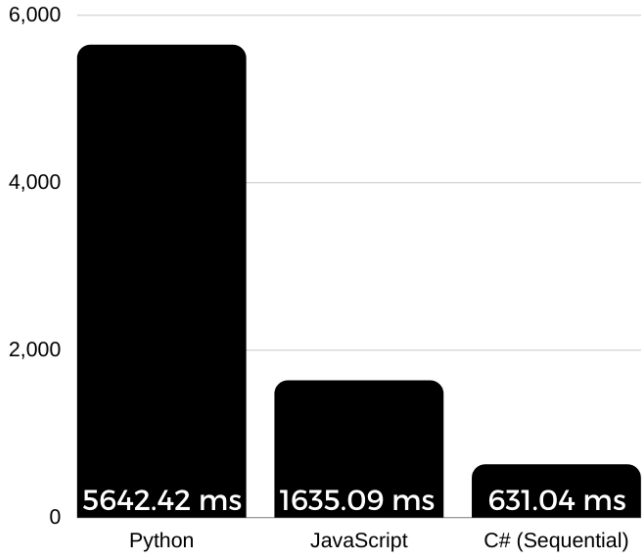

**Figure 1: Average runtime of quicksort experiment in Python, JavaScript, and C# (in milliseconds)**

Comparatively, the same operations were performed in **C#**, but twice each iteration: once the random-filled array was generated, it would be cloned as-is. One copy of the array would be sorted using a standard (sequential) quicksort algorithm, and the other would be sorted using a parallelized quicksort to exploit concurrency. The average runtime between these two can be seen in Figure 2.

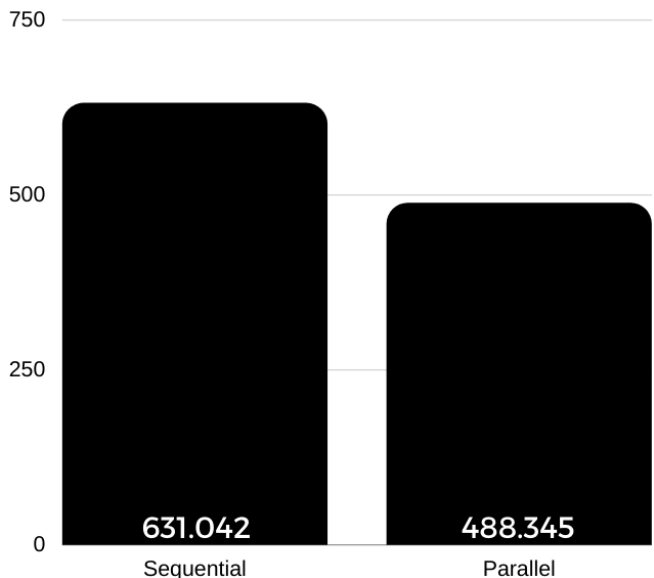

**Figure 2: Average runtime of sequential vs. parallel quicksort experiment in C# (in milliseconds)**

Figure 1 also compares Python and JavaScript to C#, one of many languages **compilable to WebAssembly** — though many other languages compile as well which also feature strong implementations of concurrency, such as C++, Rust, Golang, and Java [8].

## 4. DISCUSSION

While the performance difference between Python and JavaScript is highly evident, the near-native speed of C# still outperformed the others. There are packages such as NectarJS which can compile regular JS, among languages such as C and C++ to a heavily optimized asm.js file, or to Wasm from JS and so on [9]. This of course is potentially disastrous, as the compiled JS/Wasm files may face compatibility issues with each other due to the JavaScript interop required to load these files.

A better course of action will be to compile heavy calculation sections instead, and use plain JavaScript as a **compatibility layer** rather than compiling all files into Wasm. For example, in the algorithm mentioned in *Section 3.1*, rather than attempting to concurrently perform every operation such as populating the array, in the C# implementation we utilized concurrency only during the "calculation" section, ie. the quicksort. While not applicable in this particular case, with this same technique applied to the data cube visualization algorithms, we retain a sense of **data and type safety** while still gaining a boost in performance.

### 4.1 Application Usage

In order to utilize concurrency on the web application, we are faced with two options:
a) Rewrite the entire backend of the CubeViewer client and use something such as JS/C# interop with the Blazor WebAssembly framework or Mono-Wasm [10].

b) Rewrite the bulk of the primary cube operation functions from JS into another language and compile directly to Wasm.

The latter in this case is the better choice of the two our optimization purposes. Given the set of languages that are compilable to Wasm, the easiest choices would be either Rust or Golang, since these languages compile natively already and feature concurrency within the language as well. However, even for **serial (or sequential)** algorithms, compiling the JavaScript (through TypeScript) from the experiment algorithm mentioned in *Section 3.1* into Wasm returned the fastest overall performance out of all three languages tested in Figure 1 initially. Figure 3 shows us the result of this new experiment.

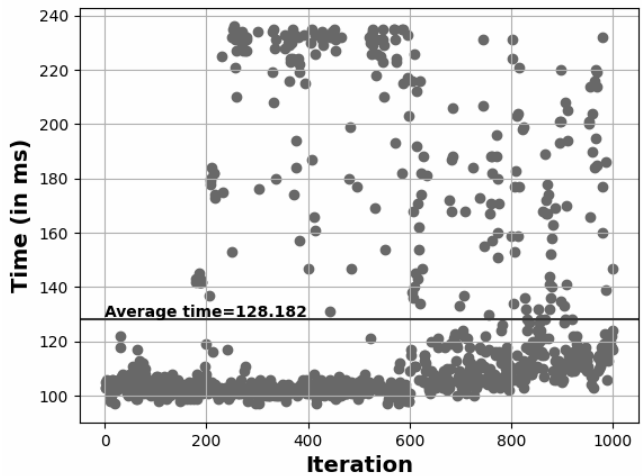

**Figure 3: Quicksort algorithm duration, JS compiled to Wasm through TypeScript**

Performing the same experiment of *Section 3.1* using Golang instead, both in serial and parallel algorithm forms, also yields interesting results, as seen in Figure 4. Using an online tool [11] to compile Golang to WebAssembly and run the code on the fly showed that WebAssembly performs nearly identical to native Golang when operating using concurrency — this will be the basis for future research.

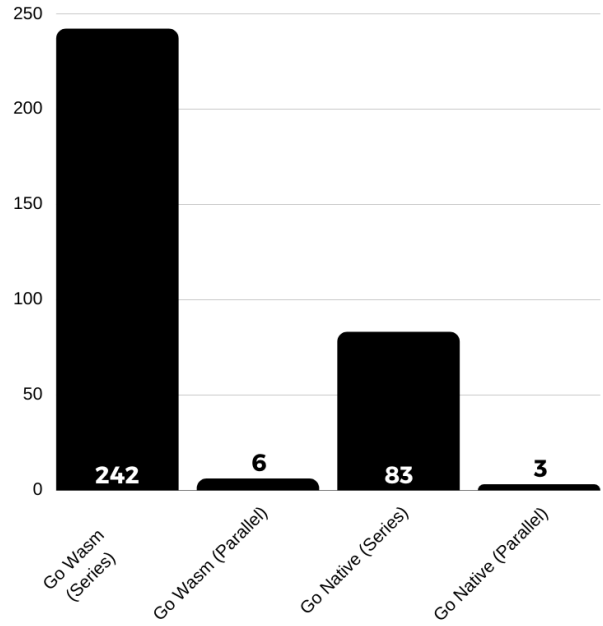

**Figure 4: Quicksort algorithm duration, Golang compiled to Wasm vs. native performance, series and parallel (measured in ms)**

## 5. CONCLUSION

This paper focused on two major problems with data cubes visualization: time and oversimplification of data. Current popular web-friendly languages such as Python and JavaScript (and most other dynamically typed languages) perform slowly and lack native support for high-precision data types.

As a result, current non-commercial visualization applications written for cross-platform usage (web, mobile) suffer greatly from these drawbacks. Through our unit testing, we were able to discover that the usage of concurrency for large data sets was overall more efficient for big data processing. This factor, combined with the recent development of WebAssembly, allows us to write explicitly-typed code with high-precision data types, while also being able to utilize shared memory concurrency as well.

Being able to solve these two issues utilizing one programming language tailored towards these issues will allow us the ability to integrate

WebAssembly into existing web applications, such as CubesViewer which performs its computations through JavaScript. By using interopable functions between WebAssembly and JavaScript, we can perform intense data cube operations such as aggregation at near-native speed while retaining data precision and minimizing the amount of code needing to be rewritten.

Future research will focus on this integration of WebAssembly into CubesViewer, utilizing programming languages such as Golang which has a thorough implementation of built-in concurrency and compiling this into our JS/Wasm interop algorithms; additionally, future research may be performed to test the usage of concurrent data structures to aid in these developments.